\documentclass[proceedings]{crckbked}   
\usepackage{graphics}

\begin{document}

\begin{article}
\begin{opening}
\title{He-D$_3$ polarization observed in prominences}
\runningtitle{D$_3$ polarization obs. in prominences}
\runningauthor{R. Ramelli, M. Bianda}
\author{R. \surname{Ramelli}}
\author{M. \surname{Bianda}}
\institute{Istituto Ricerche Solari Locarno (IRSOL)\\ 6605 Locarno Monti, Switzerland}

\begin{abstract}
Spectro-polarimetric measurements of the D$_3$-He {\sc i} 5876 \AA \ line profile in 35 
prominences have been performed in 2003 with the Gregory-Coud\'e Telescope
in Locarno. Two different experimental techniques (ZIMPOL and
Beam-exchange method) have been successfully employed to determine 
all four Stokes components. Both give compatible
results.
The preliminary results as well as the measurement techniques are reported.
\end{abstract}
\end{opening}

\section {Introduction}
The understanding of the prominence formation is strongly connected
with the knowledge of the magnetic fields which are responsible 
for their support.
Polarimetric observations of the emission lines in prominences
allow to study the structure and intensity of their magnetic fields
through the Hanle and the Zeeman effect.
After the extensive measurements of \inlinecite{Leroy84} and
\inlinecite{Athay83},
in the last
two decades the observational activity in this domain was rather poor.
On the other hand the progress in the instrumentation 
improved remarkably the polarimetric sensitivities that can now be
achieved.
Recently, new polarimetric observations of few prominences were reported
(e.g. \citeauthor{Paletou01}, \citeyear{Paletou01};
\citeauthor{Wiehr03}, \citeyear{Wiehr03}). In the present work
we proceed to a more extended observational program.

\section{Observations}

The evacuated Gregory-Coud\'e telescope which is used at IRSOL
(Istituto Ricerche Solari Locarno), has the advantage
to introduce small instrumental polarization and 
cross-talks \cite{sanchez91}. 
Both are generated mainly from two off-axis flat mirrors, whose
effects theoretically cancel out around the equinox.
Otherwise they increase almost linearly 
with declination and 
stay almost constant over one day of solar observations.
At solstice the total instrumental linear polarization 
reaches almost 3\% and 
the circular polarization is smaller than 1\%.

19 prominences were observed from 24th March to 24th April 2003 using the
beam exchange technique proposed by 
\inlinecite{Semel93}. 
This technique allows a measurement free from effects introduced by
the detector gain table. The data reduction
technique and the instrumental set-up are described by \inlinecite{bianda98}. 
For each prominence we took various sets of measurements at different
locations and times. Each set included typically about 20 exposures of 5
seconds each on a fixed position on the prominences, from which all four
Stokes components were extracted. 
 
A second observation series was performed using the Zurich Imaging
Polarimeter ZIMPOL{\sc ii} \cite{stenflo92,povel95,gandorfer97,gandorfer99},
which allows to get polarization measurements free from seeing
effects. 
16 prominences were observed from 22nd May to 26th September 2003. 
The image was rotated with a Dove prism set after the analyzer in order
to keep the limb parallel to the spectrograph slit. 
Various sets of measurements were
taken at different locations. Each set included typically 100
images of 10 seconds exposure each.

In both observing techniques additional 
measurements were regularly performed for
calibration purposes. The light originating at center of the solar
disc was assumed to be unpolarized and therefore 
used as reference to establish the correction 
for the instrumental polarization.
The background intensity profile was determined 
observing a quiet region of the halo near the
prominence.
We sometimes measured a nonzero polarization
in the background light in a particular region (NE) above the solar limb 
(values up to 7\%). 
This is believed to come from spurious reflections within the
telescope.  
The full Stokes vector of the background light
was therefore subtracted from the total measured Stokes vector.

In order to account for
the cross-talk from the linear polarization to the
circular polarization (values up to about 20\% at solstice)
measurements were performed applying
a linear polarization filter in different positions before the entrance window 
of the telescope. 
The cross-talk from the circular to the linear polarization was not 
taken into account since it was expected to be negligible.

\section{Results}

\begin{figure}
\centerline{
{\resizebox{11.6cm}{!}{\includegraphics{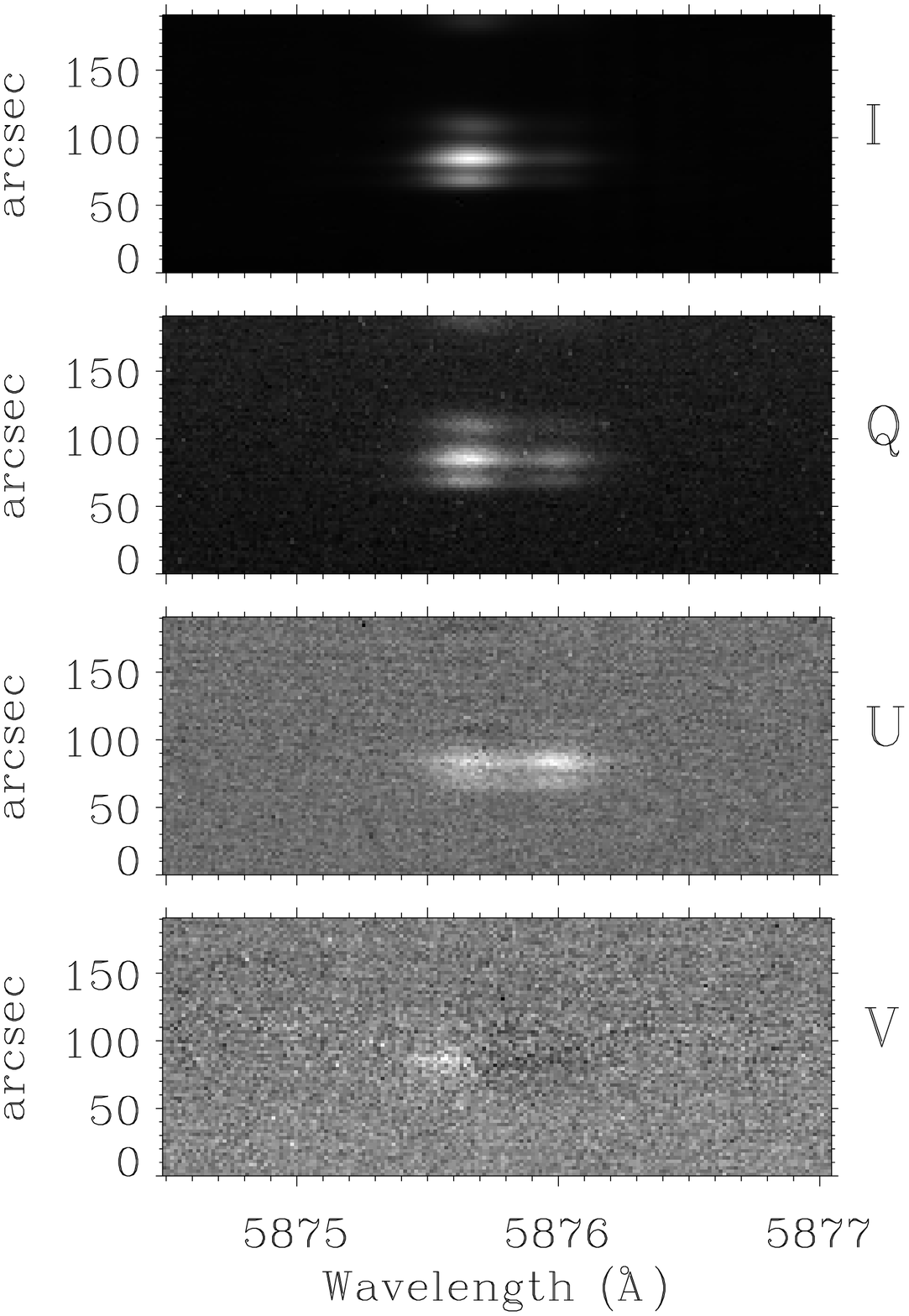}\hspace{15mm}\includegraphics{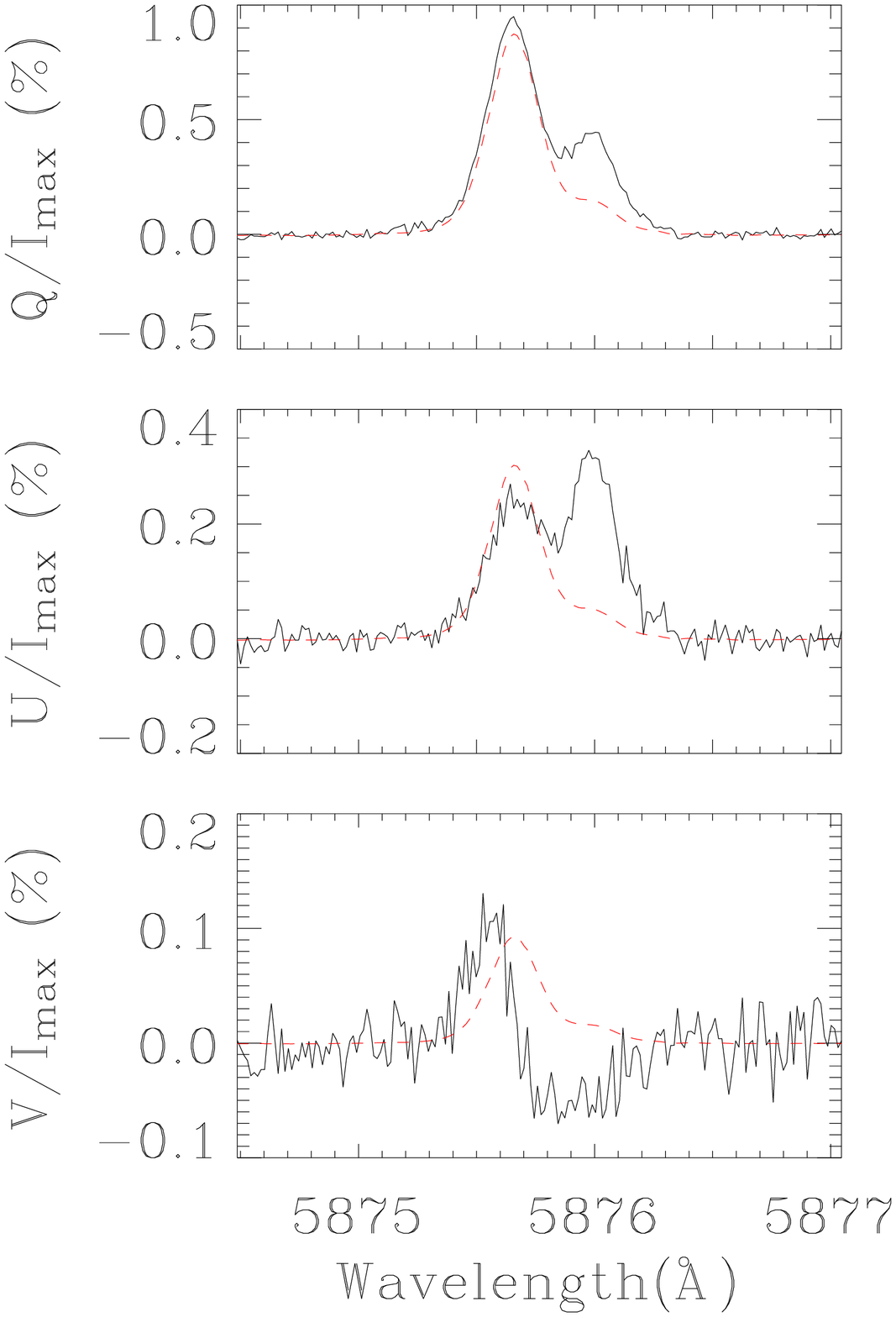}}}}
\caption{Left: Example of spectral images of the four Stokes parameters
(obtained with the ZIMPOL{\sc ii} system on the 23rd May 2003). 
Right: Polarization profiles obtained from the same images integrating
the spatial region where the intensity signal is larger than half the maximum.
Shown are Stokes $Q$, $U$
and $V$ divided by the maximum of the intensity profile $I_{\rm max}$. The scaled
intensity profile is shown by a hatched line.}
\label{stokes}
\end{figure}

\begin{figure}
\centerline{\resizebox{12cm}{!}{\includegraphics{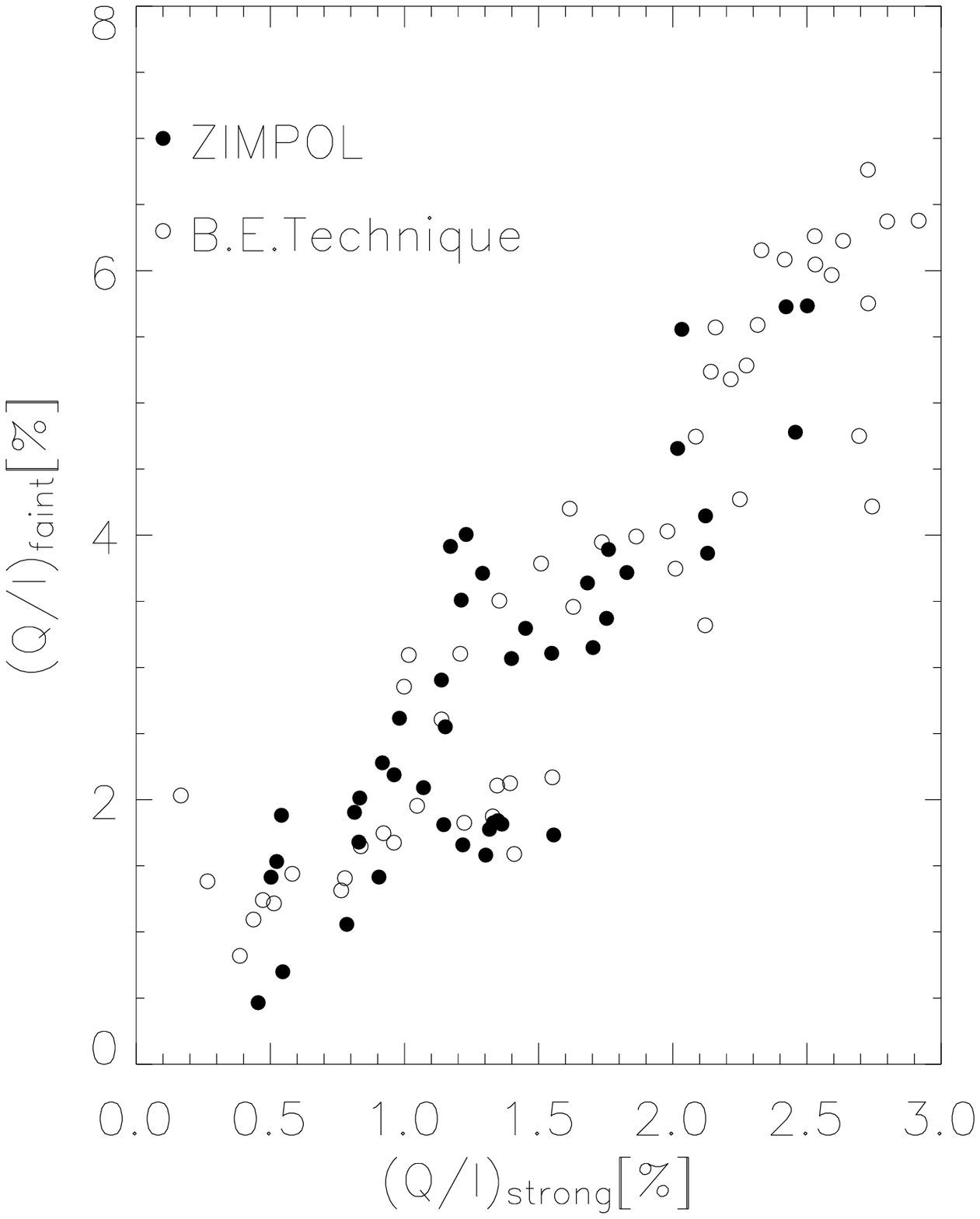}\hspace{15mm}\includegraphics{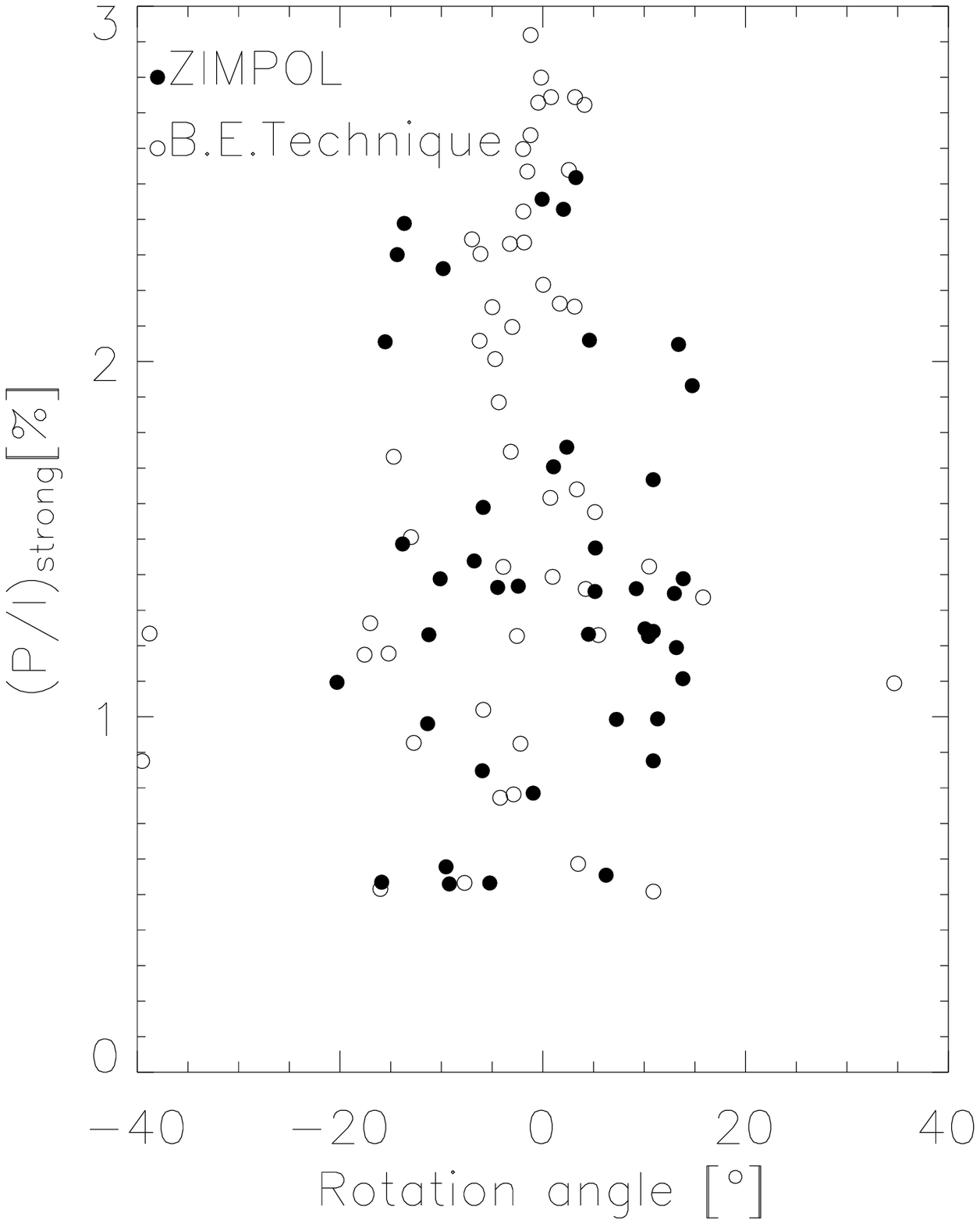}} }
\caption{Left: Scatter plot representing Stokes $Q/I$ measured in
the faint red component of the D$_3$ multiplet 
versus Stokes $Q/I$ measured in the 
strong blue component. Right: Hanle diagram showing 
the total linear polarization versus the rotation angle $\alpha=
\arctan(U/Q)$  for the strong blue component.}
\label{sumgraph}
\end{figure}

An example of the four Stokes spectral images resulting from the measurement
of one prominence with the ZIMPOL{\sc ii} system is reported
to the left in Fig. \ref{stokes}. 
In this preliminary analysis we look at the spectro-polarimetric 
profiles obtained integrating over the spatial region 
where the signal intensity is at least 50\% of the maximum intensity.
The profiles of Stokes $Q$, $U$ and $V$ divided by $I_{\rm max}$ 
(the maximum of the intensity profile)
are shown in the right panel of Fig. \ref{stokes}.
Two resolved multiplet components can be observed:
a strong blue component at 5875.6 \AA \ and a faint red component at
5875.97 \AA . To analyze the behavior of the two components 
we average the value of the relative polarization in the intervals
5875.5-5875.7 \AA\ respectively 5875.9-5876.0 \AA .
It is found that usually $Q/I$ is about a factor of two larger in the faint
component than in the strong component, as can be observed
in the left plot of Fig. \ref{sumgraph}. In the right panel we
show indeed for the strong component
the Hanle diagram, where the total linear
polarization versus the rotation angle $\alpha= \arctan(U/Q)$ is
plotted. The results obtained with the two techniques for different
prominences are similar.

\section{Conclusions}
The instrumentation at IRSOL and the
two techniques used allowed precise
measurements of the profiles of all four Stokes components
in the He{\sc i}-D$_3$ emission line of 35 prominences.
$Q/I$ was found to be always positive with values 
up to 7\% in the faint red multiplet component 
and up to 3\% in the blue component. The
absolute values of $U/I$ and $V/I$ were generally below 2\% 
and 0.5\% respectively. 

\acknowledgements
We thank Prof. Jan Stenflo and his group who allow us to use
the ZIMPOL{\sc ii} system.
We are grateful for the financial support that has been provided by
the canton of Ticino, the city of Locarno, and ETH Zurich.
We appreciated the  helpful and interesting
discussions that we had with Prof. Jan Stenflo,
Prof. Javier Trujillo Bueno and Laura Merenda.

\end{article}


\begin{thebibliography}{}

\bibitem[\protect \citeauthoryear{Athay {\it et~al.}}{1983}]{Athay83}
Athay, R.G., Querfeld, C.W., Smartt, R.N., Landi Degl'Innocenti, E., and
Bommier, V.: 1983, {\it Solar Phys.} {\bf 89}, 3.

\bibitem[\protect \citeauthoryear{Bianda {\it et~al.}}{1998}]{bianda98}
Bianda, M., Solanki, S.K., and Stenflo J.O.: 1998, {\it Astron. Astrophys.} {\bf 331}, 760.

\bibitem[\protect \citeauthoryear{Gandorfer}{1999}]{gandorfer99}
Gandorfer, A.: 1999, {\it Opt. Eng.} {\bf 38}, 1402.

\bibitem[\protect \citeauthoryear{Gandorfer and Povel}{1997}]{gandorfer97}
Gandorfer, A. and Povel, H.P.: 1997, {\it Astron. Astrophys.} {\bf 328}, 381.

\bibitem[\protect \citeauthoryear{Leroy {\it et~al.}}{1984}]{Leroy84}
Leroy, J.L., Bommier, V., and Sahal.-Br\'echot, S.: 1984, {\it Astron. Astrophys.} {\bf 131}, 33.

\bibitem[\protect \citeauthoryear{Paletou {\it et~al.}}{2001}]{Paletou01}
Paletou, F., L\'opez Ariste, A., Bommier, V., and Semel, M.: 2001, {\it Astron. Astrophys.} {\bf 375}, L39.

\bibitem[\protect \citeauthoryear{Povel}{1995}]{povel95}
Povel, H.P.: 1995, {\it Opt. Eng.} {\bf 34}, 1870.

\bibitem[\protect \citeauthoryear{Sanchez {\it et~al.}}{1991}]{sanchez91}
Sanchez Almeida, J., Mart\'inez Pillet, V., and Wittmann A.D.: 
1991, {\it Solar Phys.} {\bf 134}, 1.

\bibitem[\protect \citeauthoryear{Semel  {\it et~al.}}{1993}]{Semel93}
Semel, M., Donati, J.-F., and Rees, D.E.: 1993, {\it Astron. Astrophys.} {\bf 278}, 231.

\bibitem[\protect \citeauthoryear{Stenflo {\it et~al.}}{1992}]{stenflo92}
Stenflo, J.O., Keller, C.U., and  Povel, H.P.: 1992, LEST Foundation
Technical Report No. {\bf 54}, Univ.Oslo.

\bibitem[\protect \citeauthoryear{Wiehr and Bianda}{2003}]{Wiehr03}
Wiehr, E. and Bianda, M.: 2003, {\it Astron. Astrophys.} {\bf 404}, L25.

\end{thebibliography}
\end{document}